\documentclass{article}

\usepackage{PRIMEarxiv}

\usepackage[utf8]{inputenc} 
\usepackage[T1]{fontenc}    
\usepackage{hyperref}       
\usepackage{url}            
\usepackage{booktabs}       
\usepackage{amsfonts}       
\usepackage{amsmath}        
\usepackage{amssymb}        
\usepackage{nicefrac}       
\usepackage{microtype}      
\usepackage{lipsum}
\usepackage{fancyhdr}       
\usepackage{graphicx}       
\usepackage{tikz}           
\usepackage{pgfplots}       
\usepackage{tabularx}       
\usepackage{array}          
\usetikzlibrary{positioning}
\pgfplotsset{compat=1.16}
\graphicspath{{media/}}     

\DeclareMathOperator*{\argmax}{arg\,max}
\DeclareMathOperator{\sign}{sign}
\DeclareMathOperator{\clip}{clip}
\DeclareMathOperator{\broadcast}{broadcast}

\title{Radio Adversarial Attacks on EMG-based Gesture Recognition Networks

}

\author{
  Hongyi Xie \\
  ShanghaiTech University \\
  \texttt{xiehy@shanghaitech.edu.cn} \\
}

\begin{document}
\maketitle

\begin{abstract}
Surface electromyography (EMG) signal-based gesture recognition enables non-invasive human-computer interaction in medical rehabilitation, prosthetic control, and virtual reality. Deep learning models, such as EMGNet, achieve classification accuracies exceeding 97\%. However, these systems exhibit vulnerabilities to adversarial attacks, predominantly studied in the digital domain where perturbations are added post-collection, overlooking physical feasibility.
    
    This paper introduces ERa Attack, a radio frequency (RF) adversarial method targeting consumer-grade EMG devices like Myo Armband under intentional electromagnetic interference (IEMI). Assuming white-box access, attackers deploy low-power software-defined radio (SDR) transmitters within meters to inject optimized RF perturbations, misleading downstream models.
    
    The approach extends digital adversarial samples to the physical domain: Projected Gradient Descent (PGD) generates time-frequency perturbations against EMGNet; inverse Short-Time Fourier Transform (ISTFT) extracts components in the 50-150 Hz band; fixed frequency-domain strategies (constant spectrum noise or narrowband modulation) enable synchronization-free attacks. Perturbations, constrained to 1-10\% of signal amplitude, are amplitude-modulated onto a 433 MHz carrier and transmitted via HackRF One for electromagnetic coupling.
    
    Experiments on the Myo Dataset (7 gestures, 50 repetitions each) demonstrate efficacy: at 1 m and 0 dBm, accuracy drops from 97.8\% to 58.3\%, with a 41.7\% misclassification rate and 25.6\% attack success rate for targeted misguidance. Effects decay exponentially with distance, recovering to over 85\% at 3 m; increasing power to 10 dBm reduces accuracy by an additional 15\% at 1 m.
    
    This work pioneers RF injection in EMG recognition, enhances attack practicality via synchronization-free strategies, and quantifies perturbation modes. It underscores risks in safety-critical applications and suggests defenses like hardware shielding, spectrum monitoring, and adversarial training, informing robust EMG system design.
\end{abstract}


\section{Introduction}
\label{sec:intro}

Surface electromyography (sEMG)-based gesture recognition enables non-invasive human-computer interaction in prosthetic control and medical rehabilitation. Deep learning models, such as EMGNet, achieve classification accuracies exceeding 98\% on datasets like Myo~\cite{Chen2020HandGestureRecognition}. However, security analyses of these systems remain limited, with prior work focusing on digital-domain adversarial attacks~\cite{Kang2023SyntheticEMG}. Specifically, these attacks add perturbations to post-collection signals, neglecting vulnerabilities at the physical layer. In contrast, radio frequency (RF) injection can interfere with signal acquisition at the source, bypassing traditional encryption. Such attacks pose severe risks in safety-critical applications, including prosthetic control, where misclassification may trigger unintended actions~\cite{Leonardis2015HandExoskeleton, Cipriani2011SharedControl}.

EMG signals capture muscle activity non-invasively, facilitating applications in sports science, prosthetic manipulation, and virtual reality~\cite{Jia2020StealingMoves, Chen2020HandGestureRecognition}. Consumer-grade devices like the Myo Armband, with eight dry electrodes sampling at 200 Hz, integrate seamlessly into wearable systems for gesture-controlled drones or smart homes~\cite{Yamagata2021GeneratingAdversarialEMG, Mendez2017MyoArmbandClassification}. For instance, Cipriani et al. implemented shared prosthetic control via EMG, enhancing user naturalness~\cite{Cipriani2011SharedControl}, while Leonardis et al. developed an EMG-driven exoskeleton for bilateral stroke rehabilitation~\cite{Leonardis2015HandExoskeleton}.

Deep learning has advanced EMG gesture recognition substantially. Traditional methods rely on handcrafted features, such as root mean square (RMS) in the time domain or mean frequency (MNF) in the frequency domain, paired with classifiers like support vector machines (SVM) or k-nearest neighbors (KNN)~\cite{DeLuca2006Decomposition}. These approaches falter in large gesture sets or cross-user scenarios due to limited robustness~\cite{Atzori2015NinaproDatabase}. Convolutional neural networks (CNNs) address this by learning spatiotemporal features end-to-end. Atzori et al. applied a four-layer CNN to the NinaPro dataset, matching traditional performance without manual features~\cite{Atzori2015NinaproDatabase}. Wei et al. introduced a multi-stream CNN (MSCNN) for channel-specific feature fusion, outperforming single-stream models~\cite{Wei2019MultiStreamCNN}. Chen et al.'s compact EMGNet, using continuous wavelet transform (CWT) spectrograms as input, attains 98.8\% accuracy on Myo data with halved parameters compared to prior CNNs~\cite{Chen2020HandGestureRecognition}.

Despite these gains, EMG systems exhibit security vulnerabilities inherent to deep models. Digital-domain attacks synthesize user-specific signals via generative adversarial networks (GANs) to spoof authentication~\cite{Jia2020StealingMoves} or apply fast gradient sign method (FGSM) perturbations to time-frequency representations, reducing accuracy from near 100\% to near 0\%~\cite{Yamagata2021GeneratingAdversarialEMG, Kang2023SyntheticEMG}. However, these assume access to digitized signals, impractical in real-time scenarios without compromising transmission links.

Intentional electromagnetic interference (IEMI) offers a physical-layer alternative, injecting RF signals remotely to disrupt electronics~\cite{Radasky2004IEMI, Radasky2010IEMIImpact, Kasmi2015IEMIThreats}. IEMI exploits front-door (e.g., antennas) or back-door (e.g., cables, seams) coupling paths~\cite{Radasky2004IEMI}. Wearable devices like Myo, prioritizing compactness and cost, often lack robust shielding, making them susceptible to back-door attacks via electrode lines acting as unintended antennas~\cite{DeLuca2006Decomposition}.

The Brain-Hack attack exemplifies this on EEG systems, using software-defined radio (SDR) to amplitude-modulate (AM) low-frequency signals onto a 500 MHz carrier, exploiting amplifier nonlinearity for demodulation and injection of false brainwaves~\cite{ArmengolUrpi2023BrainHack}. This enables remote control of brain-computer interfaces (BCIs), such as inducing drone crashes or falsifying stress data.

Adapting Brain-Hack to EMG faces a core challenge: sEMG amplitudes (millivolt-level) exceed EEG (microvolt-level) by 2--3 orders of magnitude~\cite{DeLuca2006Decomposition}. Overwhelming sEMG requires 40--60 dB higher power ($P \propto V^2$), rendering it infeasible. Prior EMG attacks, limited to digital simulations~\cite{Kang2023SyntheticEMG, Xue2022UniversalEMG}, overlook this physical constraint.

This paper addresses the gap by proposing ERa Attack, a RF adversarial method for EMG gesture recognition. ERa Attack shifts from signal overwhelming to model-informed deception: optimized perturbations, 1--10\% of sEMG amplitude, exploit amplifier nonlinearity and model gradients to mislead classification at low power. This leverages non-linear demodulation to inject perturbations without dominating the signal, ensuring feasibility for millivolt-level biosignals.

The contributions are as follows:

\begin{itemize}
    \item We introduce the first RF adversarial attack on EMG gesture recognition, combining adversarial sample generation with RF injection to enable remote, non-contact interference under a white-box threat model.
    \item We design a white-box optimization for RF perturbations, including the EMI-FGSM algorithm, which enforces channel-consistent gradients to align with physical interference, achieving higher efficacy at lower power than random noise or fixed modulations~\cite{ArmengolUrpi2023BrainHack}.
    \item We construct a low-cost HackRF One-based platform and validate ERa Attack on Myo Armband with the Myo dataset, quantifying impacts of distance, power, and perturbation modes on accuracy (e.g., dropping from 97.8\% to 58.3\% at 1 m, 0 dBm).
    \item We propose multi-layer defenses, spanning hardware shielding, signal anomaly detection, and adversarial training, to mitigate such attacks.
\end{itemize}

The remainder of the paper is organized as follows. Chapter 2 reviews theoretical foundations, including EMG characteristics, IEMI principles, and adversarial examples. Chapter 3 defines the problem and threat model. Chapter 4 details ERa Attack's architecture. Chapter 5 presents experimental setup and results. Chapter 6 concludes with limitations and future directions.

\section{Background and Related Work}
\label{sec:relwork}
Surface electromyography (sEMG) signals arise from muscle activity, enabling non-invasive gesture recognition in applications such as prosthetic control. Deep learning models process these signals, yet vulnerabilities to physical-layer attacks remain underexplored. This chapter outlines the physiological and engineering aspects of sEMG, details representative acquisition hardware and recognition models, explains intentional electromagnetic interference (IEMI) mechanisms, and reviews adversarial attack principles. Prior work on biosignal security focuses on digital perturbations or EEG-specific injections, overlooking millivolt-level sEMG amplitudes that necessitate low-power, model-informed disturbances. In contrast, our approach optimizes perturbations for EMGNet via gradient-based methods, ensuring efficacy under physical transmission constraints.

\subsection{Surface Electromyography Signals}

sEMG signals manifest as electrical potentials on the skin surface during muscle contractions, governed by motor unit (MU) dynamics. Each MU comprises an alpha motoneuron and its innervated muscle fibers, activating synchronously under the all-or-none principle~\cite{DeLuca2006Decomposition}. Neural impulses trigger motor unit action potentials (MUAPs), which are triphasic pulses with peak-to-peak amplitudes of approximately 0.5 mV and durations of 8--14 ms~\cite{DeLuca2006Decomposition}. Muscle force modulation occurs via spatial recruitment, activating progressively larger MUs, and temporal rate coding, increasing firing rates from 5 Hz at onset~\cite{DeLuca2006Decomposition}.

Recorded sEMG represents the spatiotemporal superposition of numerous MUAPs, forming an interference pattern characterized as stochastic and non-stationary~\cite{DeLuca2006Decomposition}. Amplitudes range from 0--10 mV peak-to-peak or 0--1.5 mV root mean square (RMS), exceeding EEG by 2--3 orders of magnitude~\cite{DeLuca2006Decomposition}. Spectral energy concentrates in 20--500 Hz, with dominant contributions at 50--150 Hz; fatigue shifts the spectrum toward lower frequencies~\cite{DeLuca2006Decomposition}.

These characteristics inform attack design: perturbations must align with the 50--100 Hz band for Myo Armband's 200 Hz sampling to avoid aliasing while targeting model-sensitive features. Unlike EEG attacks that overwhelm microvolt signals, sEMG requires perturbations at 1--10

\subsection{EMG Acquisition and Recognition Systems}

Consumer-grade devices like Myo Armband acquire sEMG via eight dry stainless steel electrodes, sampling at 200 Hz with 8-bit resolution~\cite{Mendez2017MyoArmbandClassification}. An ARM Cortex M4 processor handles initial processing, while a 9-axis inertial measurement unit (IMU) captures motion. Data transmits via Bluetooth Low Energy (BLE) at 2.4 GHz. Interconnecting flexible printed circuit boards (PCBs), spanning 19--34 cm without shielding, act as unintentional antennas for back-door coupling~\cite{Chen2020HandGestureRecognition}.

EMGNet, a lightweight CNN, processes these signals for gesture classification~\cite{Chen2020HandGestureRecognition}. Input preprocessing applies continuous wavelet transform (CWT) to 52-sample windows across eight channels, yielding 8$\times$15$\times$25 tensors after downsampling. The architecture employs four 3$\times$3 convolutional layers interspersed with max pooling, culminating in global average pooling without fully connected layers, achieving 98.8

Prior EMG recognition relies on handcrafted features (e.g., RMS, median frequency) and classifiers like SVM, limited in cross-user scenarios~\cite{DeLuca2006Decomposition}. Deep models like EMGNet surpass these, yet expose vulnerabilities. Digital attacks synthesize perturbations via GANs or FGSM, dropping accuracy from near 100

\subsection{Intentional Electromagnetic Interference}

IEMI entails deliberate electromagnetic emissions to disrupt electronics, categorized by coupling paths~\cite{Radasky2004IEMI}. Front-door attacks target designed ports like antennas; back-door attacks exploit unintended paths such as cables or seams~\cite{Radasky2004IEMI}. Wearables like Myo, lacking robust shielding, are susceptible to back-door injections via flexible PCBs~\cite{Radasky2004IEMI}.

Amplifier nonlinearity enables demodulation of amplitude\-modulated (AM) signals. Real amplifiers exhibit Taylor series responses:
\begin{equation}
X_{out} = A_1 X_{in} + A_2 X_{in}^2 + \cdots
\end{equation}
For AM input $s(t) = A_c [1 + k_a m(t)] \cos(2\pi f_c t)$, the quadratic term yields low-frequency components recovering $m(t)$ after filtering high harmonics~\cite{Radasky2004IEMI}. Brain-Hack exploits this on EEG, injecting via 500 MHz carriers to overwhelm microvolt signals~\cite{ArmengolUrpi2023BrainHack}. However, sEMG's millivolt amplitudes demand 40--60 dB higher power for overwhelming ($P \propto V^2$), rendering direct adaptation infeasible. Our method injects optimized perturbations at low amplitudes, leveraging nonlinearity for superposition rather than dominance.

\subsection{Adversarial Attacks on Deep Learning Models}

Adversarial examples perturb inputs to induce misclassification, formulated as minimizing $|\delta|_p$ subject to $f(x + \delta) = y_{target}$~\cite{Goodfellow2015ExplainingHarnessing}. Equivalently, maximize loss $J(\theta, x + \delta, y)$ with $|\delta|_p \leq \epsilon$. Fast Gradient Sign Method (FGSM) computes $\delta = \epsilon \cdot \sign(\nabla_x J(\theta, x, y))$ in one step~\cite{Goodfellow2015ExplainingHarnessing}. Projected Gradient Descent (PGD) iterates: $x^{t+1} = \Pi_\epsilon(x^t + \alpha \cdot \sign(\nabla_x J(\theta, x^t, y)))$, yielding stronger perturbations~\cite{Madry2018TowardsDeepLearning}.

Physical attacks bridge sim-to-real gaps via Expectation Over Transformation (EOT), optimizing $\argmax_\delta \mathbb{E}_{t \sim \mathcal{T}} [J(\theta, t(x + \delta), y)]$ to robustify against distortions like attenuation or noise~\cite{athalye2018synthesizing}. Digital biosignal attacks apply FGSM/PGD to EEG/EMG spectrograms, reducing accuracy substantially~\cite{Goodfellow2015ExplainingHarnessing, Madry2018TowardsDeepLearning}, but neglect physical injection. We extend EOT to model IEMI channels, generating perturbations robust to distance (1--3 m decay) and sampling limits, distinguishing from overwhelming strategies by emphasizing low-power deception.

\section{Methodology}
\label{sec:methodology}

\subsection{Threat Model and Goals}
\label{chap:threat_model}

We formalize the threat model for ERa Attack, a radio frequency (RF) adversarial injection targeting surface electromyography (sEMG) gesture recognition systems. The model assumes a white-box adversary with access to the target deep learning classifier, such as EMGNet, and leverages intentional electromagnetic interference (IEMI) to inject perturbations at the signal acquisition stage. This section delineates the attack scenario, adversary capabilities, assumptions, and goals, while contrasting with prior work like Brain-Hack~\cite{ArmengolUrpi2023BrainHack}.

\subsubsection{Attack Scenario and Assumptions}

The attack unfolds in an indoor environment where a victim wears a consumer-grade sEMG device, such as Myo Armband, to control applications via gestures (e.g., fist, open palm). The adversary, positioned within meters (e.g., 1--3 m), deploys a low-cost software-defined radio (SDR) setup, like HackRF One connected to a portable computer running GNU Radio, to emit optimized RF signals. These signals couple into the device's analog front-end via back-door paths, superimposing adversarial perturbations on raw sEMG signals before digitization, thereby misleading the downstream EMGNet model~\cite{Chen2020HandGestureRecognition}.

This scenario relies on the following assumptions:

\begin{itemize}
    \item \textbf{Device Vulnerability:} Consumer-grade sEMG devices lack robust electromagnetic shielding due to cost and wearability constraints. Myo Armband's unshielded flexible printed circuit board (Flex-PCB), spanning 19--34 cm, acts as an unintentional antenna for RF signals in the 400--900 MHz band, enabling back-door coupling into the analog amplifiers~\cite{ArmengolUrpi2023BrainHack}.
    \item \textbf{Proximity:} The adversary operates within a few meters of the victim, ensuring sufficient field strength at low transmit power (e.g., 0--10 dBm) for effective injection while maintaining stealth.
    \item \textbf{Channel Conditions:} The attack occurs in typical indoor multipath fading environments with background noise, requiring perturbation robustness.
    \item \textbf{Adversary Capabilities:} The adversary possesses SDR hardware for signal generation and white-box knowledge of EMGNet, including architecture, parameters $\theta$, and preprocessing (e.g., continuous wavelet transform). This enables gradient-based optimization of perturbations.
\end{itemize}

The attack chain is modeled as:
\begin{align}
x_{ADC}(t) &= H_{ADC} \Big( H_{Amp} \Big( s_{EMG}(t) + n_{env}(t) \nonumber \\
&\quad + H_{Ant} \circ H_{Prop}(s_{RF}(t, \delta_{adv})) \Big) \Big),
\label{eq:signal_model}
\end{align}
where $s_{EMG}(t)$ is the clean sEMG signal, $n_{env}(t)$ environmental noise, $\delta_{adv}$ the digital perturbation, $s_{RF}(t, \delta_{adv})$ the modulated RF signal, $H_{Prop}$ propagation, $H_{Ant}$ antenna coupling, $H_{Amp}$ nonlinear amplification (demodulating low-frequency perturbations), and $H_{ADC}$ sampling/quantization.

\subsubsection{Adversary Goals and Constraints}

The adversary aims to mislead the classifier $f(\cdot; \theta)$ under two objectives:

\begin{itemize}
    \item \textbf{Untargeted Attack:} Induce misclassification, i.e., $f(x') \neq y_{true}$ for perturbed input $x' = x + \delta_{adv}$, reducing overall accuracy (e.g., from 97.8\% to below 60\% in experiments).
    \item \textbf{Targeted Attack:} Force classification to a specific erroneous label $y_{target} \neq y_{true}$, i.e., $f(x') = y_{target}$, with success rates up to 25.6\% at 1 m and 0 dBm.
\end{itemize}

Constraints ensure feasibility and stealth:

\begin{itemize}
    \item \textbf{Power Constraint:} Transmit power $P_{tx} \leq P_{max}$ (e.g., 10 dBm) to avoid detection and comply with hardware limits.
    \item \textbf{Perturbation Budget:} $\|\delta_{adv}\|_\infty \leq \epsilon$ (e.g., 1--10\% of sEMG amplitude) for imperceptibility, measured post-demodulation via injection SNR:
    \begin{equation}
    SNR_{inj} = 10 \log_{10} \left( \frac{P_{s_{EMG}}}{P_{\delta'_{adv}}} \right),
    \label{eq:snr}
    \end{equation}
    where $P_{\delta'_{adv}}$ is the demodulated perturbation power.
\end{itemize}

Compared to Brain-Hack~\cite{ArmengolUrpi2023BrainHack}, which overwhelms microvolt-level EEG signals via black-box fixed modulations, ERa Attack targets millivolt-level sEMG with white-box gradient optimization for superposition, not drowning, achieving efficacy at 40--60 dB lower power due to $P \propto V^2$ scaling.

\subsubsection{Security Goals}

This work evaluates sEMG system vulnerabilities under the defined threat model to inform robust designs. Specifically, it quantifies attack success rates (ASR) and injection SNR to establish baselines for defenses, such as hardware shielding and adversarial training, aiming to maintain classification accuracy above 85\% even at 1 m attack distance and 10 dBm power.

\subsection{Overall Architecture}
\label{sec:era_attack_architecture}

ERa Attack integrates adversarial perturbation optimization in the digital domain with radio frequency (RF) injection in the physical domain to mislead surface electromyography (sEMG) gesture recognition models. The method addresses the challenge of bridging abstract mathematical optimizations to practical signal engineering, enabling remote, non-contact interference under the threat model defined in ~\ref{chap:threat_model}. This chapter delineates the attack's dual-stage architecture, perturbation generation algorithms, RF signal mapping, and feasibility considerations.

ERa Attack comprises two stages: offline perturbation optimization and online physical injection. The offline stage generates a low-frequency digital perturbation $\delta_{adv}$ targeting EMGNet~\cite{Chen2020HandGestureRecognition}, leveraging white-box access to model parameters $\theta$. The online stage modulates $\delta_{adv}$ onto an RF carrier for transmission via software-defined radio (SDR), exploiting amplifier nonlinearity for superposition onto raw sEMG signals.

In the offline stage, a representative clean sEMG sample $x$ undergoes iterative gradient-based optimization to maximize classification loss $\mathcal{L}(f(x + \delta_{adv}; \theta), y)$, yielding $\delta_{adv}$ with $\|\delta_{adv}\|_\infty \leq \epsilon = 8/255$. This computation-intensive process, executed once, produces a reusable perturbation template.

The online stage converts $\delta_{adv}$ to an amplitude-modulated (AM) RF signal $s_{RF}(t, \delta_{adv}) = A_c [1 + k_a \delta_{adv}(t)] \cos(2\pi f_c t)$, transmitted continuously via HackRF One. Propagation $H_{Prop}$, antenna coupling $H_{Ant}$, and nonlinear amplification $H_{Amp}$ demodulate $\delta_{adv}$, adding it to $s_{EMG}(t)$ before analog-to-digital conversion (ADC)~\cite{ArmengolUrpi2023BrainHack}. The perturbed input $x'$ induces erroneous outputs from EMGNet.

\begin{figure*}[htb]
   \centering
   \begin{tikzpicture}[
       scale=0.7,
       node distance=1.1cm, 
       auto,
       block/.style={rectangle, draw, fill=gray!10, text width=2.0cm, text centered, rounded corners, minimum height=0.9cm, font=\scriptsize}
   ]
     \node (input) [block] {Clean EMG Signal Samples $x$};
     \node (model) [block, below of=input] {EMGNet Model\\(Known Parameters $\theta$)};
     \node (gradient) [block, below of=model] {Backpropagation\\Compute Gradient $\nabla_x \mathcal{L}$};
     \node (optimizer) [block, below of=gradient] {Adversarial Perturbation\\Optimization Algorithm\\(e.g., FC-PGD)};
     \node (output1) [block, below of=optimizer] {Digital Adversarial\\Perturbation $\delta_{adv}$};

     \node (input2) [block, right of=input, xshift=6cm] {Digital Adversarial\\Perturbation $\delta_{adv}$};
     \node (modulator) [block, below of=input2] {GNU Radio\\Modulator (AM)};
     \node (transmitter) [block, below of=modulator] {HackRF One\\Transmitter};
     \node (channel) [block, below of=transmitter] {Transmit RF Signal\\$s_{RF}(t, \delta_{adv})$};
     \node (process) [block, below of=channel] {Physical Process:\\Wireless Channel\\$\to$ Myo Armband};
     \node (result) [block, below of=process] {Contaminated Signal $x'$\\Input to EMGNet};
     \node (output2) [block, below of=result] {Incorrect Classification\\Result $y_{adv}$};

     \draw[->] (input) -- (model);
     \draw[->] (model) -- (gradient);
     \draw[->] (gradient) -- (optimizer);
     \draw[->] (optimizer) -- (output1);
     \draw[->, dashed] (output1.east) -- ++(1cm,0) -- ++(0,4cm) -- (input2.west);
     \draw[->] (input2) -- (modulator);
     \draw[->] (modulator) -- (transmitter);
     \draw[->] (transmitter) -- (channel);
     \draw[->] (channel) -- (process);
     \draw[->] (process) -- (result);
     \draw[->] (result) -- (output2);
     
     \node[above=0.5cm of input, font=\small\bfseries] {Phase 1: Offline Optimization};
     \node[above=0.5cm of input2, font=\small\bfseries] {Phase 2: Online Injection};
   \end{tikzpicture}
   \caption{Overall Architecture of the ERa Attack}
   \label{fig:era_attack_architecture}
\end{figure*}
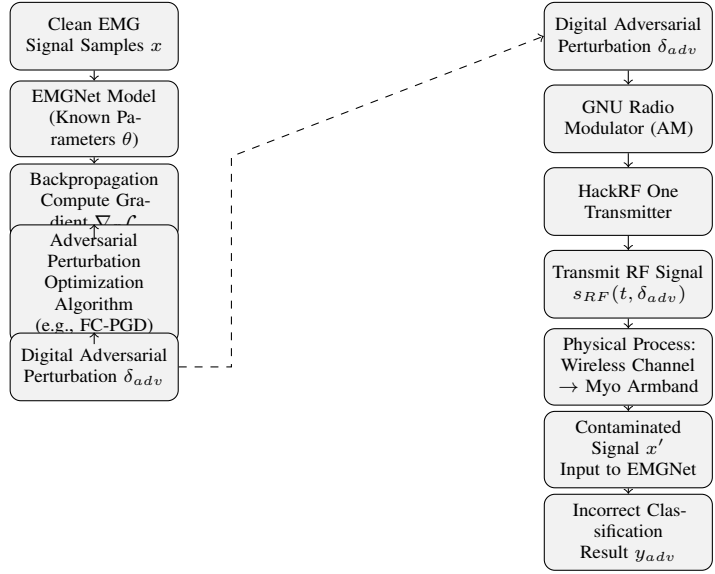

This architecture decouples complex optimization from simple emission, enhancing deployability. Figure~\ref{fig:era_attack_architecture} illustrates the workflow.

\subsection{Adversarial Perturbation Optimization}
\label{sec:adversarial_perturbation_optimization}

Perturbation optimization solves $\max_{\delta_{adv}} \mathcal{L}(f(x + \delta_{adv}; \theta), y)$ subject to $\|\delta_{adv}\|_p \leq \epsilon$, using projected gradient descent (PGD) as the base~\cite{Madry2018TowardsDeepLearning}. The algorithm employs standard hyperparameters optimized for EMGNet's time-frequency inputs, including 20 iterations with step size $\alpha = 2/255$ and perturbation budget $\epsilon = 8/255$.

\subsubsection{Time-Domain Pulse Perturbation}
\label{sec:time_domain_perturbation}

Time-domain optimization treats a 52-sample sEMG window as a vector, applying PGD directly:
\begin{equation}
\delta_{k+1} = \clip_{\epsilon} \left( \delta_k + \alpha \cdot \sign(\nabla_x \mathcal{L}(f(x+\delta_k; \theta), y)) \right).
\label{eq:time_domain_pgd}
\end{equation}
Initialization uses uniform randomness in $[-\epsilon, \epsilon]$. However, this approach falters in physical settings due to synchronization issues: perturbations must align precisely with muscle activations, challenging for remote attacks without timing knowledge~\cite{Yuan2018CommanderSong, Li2020AdvPulse}. Misalignment reduces efficacy, prompting frequency-domain alternatives for time-invariance.

\subsubsection{Frequency-Domain Perturbation (FC-PGD)}
\label{sec:frequency_domain_perturbation}

Frequency-constrained PGD (FC-PGD) generates time-invariant perturbations by averaging gradients over time:
\begin{equation}
g_{freq}[c, f] = \frac{1}{T} \sum_{t=1}^{T} g[c, f, t],
\label{eq:freq_gradient}
\end{equation}
where $g[c, f, t]$ is the gradient at channel $c$, frequency $f$, and time $t$; $T=52$. Broadcasting $g_{freq}$ yields:
\begin{equation}
\delta_{k+1} = \clip_{\epsilon} \left( \delta_k + \alpha \cdot \sign(\broadcast(g_{freq})) \right).
\label{eq:freq_pgd_update}
\end{equation}
Channel consistency averages over channels, ensuring uniform interference across Myo's eight electrodes. The resultant $\delta_{adv}$ converts to multi-tone signals, enabling continuous emission without synchronization, targeting EMGNet's spectral sensitivities~\cite{Ilyas2019AdversarialFeatures}.

\subsubsection{Hybrid Perturbation Optimization}
\label{sec:hybrid_perturbation}

Hybrid optimization combines dominant frequency-domain background $\delta_{freq}$ (via FC-PGD with $\epsilon_{freq}$) and auxiliary time-domain pulses $\delta_{time}$ (via standard PGD on $x + \delta_{freq}$ with $\epsilon_{time} < \epsilon_{freq}$), yielding $\delta_{hybrid} = \delta_{freq} + \delta_{time}$. Frequency components provide robust baseline disruption, while time pulses exploit transient features, enhancing attack strength by 10--15 percentage points in accuracy reduction under partial synchronization.

\subsection{RF Signal Mapping and Generation}
\label{sec:rf_signal_mapping_generation}

Mapping $\delta_{adv}$ to transmittable RF involves carrier selection and SDR implementation.

\subsubsection{Carrier Frequency Selection and Bandpass Design}
\label{sec:carrier_frequency_selection}

Carrier $f_c$ maximizes coupling into Myo's Flex-PCB, modeled as a half-wave dipole with length $L=19$--34 cm, yielding theoretical resonances at 441--789 MHz. Dielectric loading from human tissue shifts resonances lower, to approximately 433 MHz~\cite{ArmengolUrpi2023BrainHack}. Scanning 400--900 MHz identifies optima. A bandpass filter post-amplifier suppresses harmonics, concentrating power and minimizing interference.

\subsubsection{GNU Radio Transmission}
\label{sec:gnuradio_hackrf_implementation}

\begin{figure}[htbp]
   \centering
        \includegraphics[width=0.9\columnwidth]{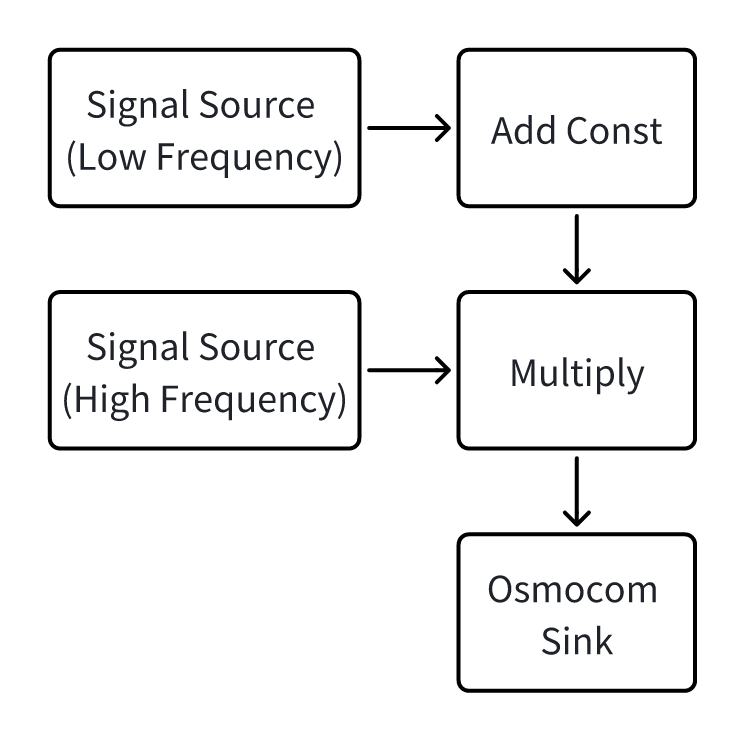}
   \caption{GNU Radio Flowgraph for the ERa Attack}
   \label{fig:gnuradio_flowgraph}
\end{figure}

Hardware chains a Linux computer, HackRF One, bandpass filter, optional amplifier (1--5 W), and antenna. GNU Radio flowgraph (Figure~\ref{fig:gnuradio_flowgraph}) generates $\delta_{adv}(t)$ via Signal Source, adds DC bias for AM envelope $1 + m \cdot \delta_{adv}(t)$ ($m$ modulation index), multiplies with carrier $\cos(2\pi f_c t)$, and streams to Osmocom Sink for transmission at 10 MS/s and 20 dB gain.

\vspace{1em}

\subsection{Feasibility Analysis and Challenges}
\label{sec:feasibility_and_challenges}

ERa Attack's feasibility hinges on information deception rather than power overwhelming. Nonlinear demodulation scales injected voltage with RF envelope squared, enabling millivolt perturbations at 1--5 W transmit power, 40--60 dB below EEG drowning requirements ($P \propto V^2$)~\cite{ArmengolUrpi2023BrainHack}.

Challenges include sim-to-real gaps from unmodeled analog filters and ADC jitter, mitigated via expectation over transformation (EOT) incorporating noise and variability~\cite{athalye2018synthesizing}. Distance sensitivity follows inverse-square decay, limiting range to 3 m at 10 dBm; user variability in arm circumference necessitates adaptive scanning. Hybrid perturbations balance robustness, achieving 58.3\% accuracy at 1 m versus 97.8\% baseline.

\subsection{Summary}

This methodology bridges digital adversarial optimization with physical RF injection through FC-PGD for time-invariant perturbations, addressing synchronization via spectral focus. RF mapping via GNU Radio enables deployment, with analyses confirming feasibility at low power despite challenges.

\section{Evaluation}
\label{sec:eval}

\label{chap:implementation_and_evaluation}

ERa Attack realizes a physical-layer adversarial injection against sEMG gesture recognition via a low-cost SDR platform. This chapter details the prototype implementation, experimental setup, and quantitative assessment. Evaluations measure classification degradation, attack success rates, and sensitivity to physical parameters like distance and power, using Myo Dataset with 7 gestures across 50 repetitions per condition.

\subsection{System Implementation}

The prototype integrates HackRF One for RF transmission with Myo Armband as the target device, bridging digital perturbation optimization to physical injection.

\subsubsection{Hardware Platform}

HackRF One, an open-source SDR, serves as the attack transmitter~\cite{Jia2020StealingMoves, Trippel2017WALNUT, ArmengolUrpi2023BrainHack}. It operates from 1 MHz to 6 GHz at up to 20 MSPS with 8-bit I/Q resolution, enabling programmable signal generation. Transmission gain is software-configurable, calibrated to output powers from -10 dBm (TX Gain 0) to 10 dBm (TX Gain 30) via Keysight E4417A power meter. An SL 10 mini log-periodic antenna (7 dBi gain at 433 MHz) connects via a bandpass filter to suppress harmonics, ensuring spectral containment.

Myo Armband collects sEMG via eight dry electrodes at 200 Hz, 8-bit resolution~\cite{Mendez2017MyoArmbandClassification, DeLuca2006Decomposition}. Its unshielded flexible PCB (19--34 cm) acts as an unintentional antenna for back-door coupling~\cite{Chen2020HandGestureRecognition}. Data streams via Bluetooth LE to a receiver PC for preprocessing and classification with EMGNet.

A portable Ubuntu 20.04 laptop controls HackRF One via USB, running GNU Radio for signal modulation and Python/PyTorch for perturbation generation.

\subsubsection{Software Stack and Workflow}

\begin{figure*}[htb]
    \centering
    \begin{tikzpicture}[
        scale=0.8,
        node distance=1.1cm, 
        every node/.style={draw, rectangle, minimum width=2.2cm, minimum height=0.9cm, align=center, font=\scriptsize}
    ]
        \node (start) {Clean EMG Signal Samples $x$};
        \node (model) [below of=start] {EMGNet Model\\(Known Parameters $\theta$)};
        \node (gradient) [below of=model] {Backpropagation\\Compute Gradient $\nabla_x \mathcal{L}$};
        \node (optimizer) [below of=gradient] {Adversarial Perturbation\\Optimization Algorithm\\(e.g., FC-PGD)};
        \node (output1) [below of=optimizer] {Digital Adversarial\\Perturbation $\delta_{adv}$};

        \node (input2) [right of=output1, xshift=5cm] {Digital Adversarial\\Perturbation $\delta_{adv}$};
        \node (modulator) [below of=input2] {GNU Radio\\Modulator (AM)};
        \node (transmitter) [below of=modulator] {HackRF One\\Transmitter};
        \node (channel) [below of=transmitter] {Transmit RF Signal\\$s_{RF}(t, \delta_{adv})$};
        \node (process) [below of=channel] {Physical Process:\\Wireless Channel\\$\to$ Myo Armband};
        \node (result) [below of=process] {Contaminated Signal $x'$\\Input to EMGNet};
        \node (output2) [below of=result] {Incorrect Classification\\Result $y_{adv}$};

        \draw[->] (start) -- (model);
        \draw[->] (model) -- (gradient);
        \draw[->] (gradient) -- (optimizer);
        \draw[->] (optimizer) -- (output1);
        \draw[->, dashed] (output1) -- (input2);
        \draw[->] (input2) -- (modulator);
        \draw[->] (modulator) -- (transmitter);
        \draw[->] (transmitter) -- (channel);
        \draw[->] (channel) -- (process);
        \draw[->] (process) -- (result);
        \draw[->] (result) -- (output2);
        
        \node[above=0.5cm of start, font=\small\bfseries, draw=none] {Phase 1: Offline Optimization};
        \node[above=0.5cm of input2, font=\small\bfseries, draw=none] {Phase 2: Online Injection};
    \end{tikzpicture}
    \caption{End-to-end experimental workflow of the ERa Attack}
    \label{fig:experimental_workflow}
\end{figure*}
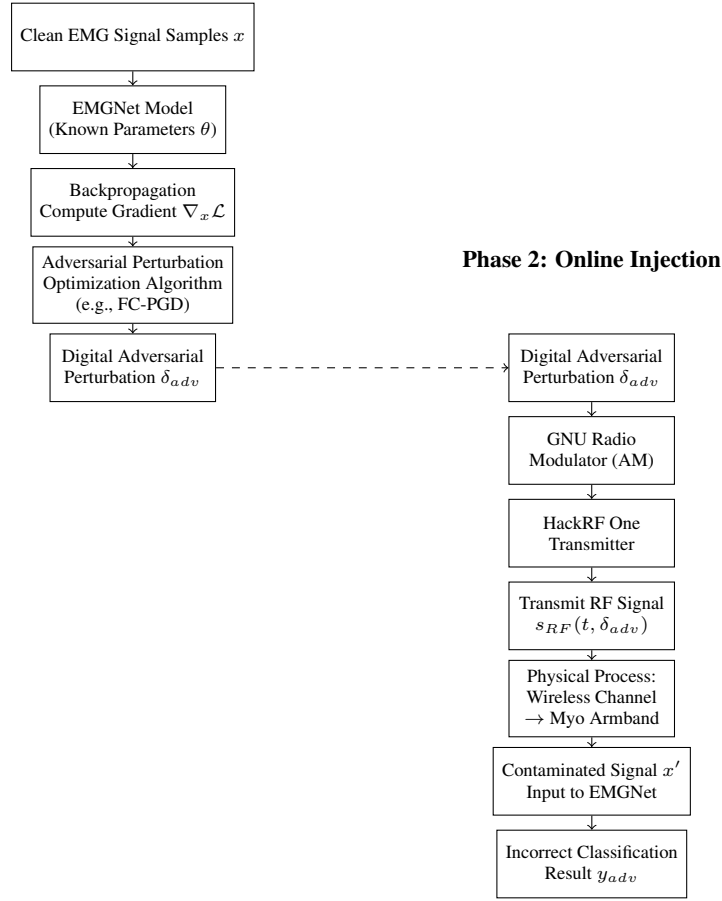

The end-to-end workflow (Figure~\ref{fig:experimental_workflow}) comprises offline optimization and online injection.

Offline: Select clean sEMG sample $x_{clean}$ from test set with label $y_{true}$. Compute $\delta_{adv}$ via PGD to maximize $\mathcal{L}(f(x_{clean} + \delta_{adv}; \theta), y)$, constrained by $\|\delta_{adv}\|_\infty \leq 8/255$.

Online: Load $\delta_{adv}$ into GNU Radio flowgraph (Figure~\ref{fig:gnuradio_flowgraph}) for AM onto 433 MHz carrier: $s_{RF}(t) = (1 + m \cdot \delta_{adv}(t)) \cos(2\pi f_c t)$, with $m=0.1$--0.3. Transmit continuously at configured power.

Victim performs gesture, inducing $s_{EMG}(t)$. RF couples, demodulates via nonlinearity, yielding perturbed $x'$. EMGNet classifies $x'$, logging $y_{pred}$ versus $y_{true}$.

Fixed seeds [42, 123, 456, 789, 999] ensure reproducibility, with 5 repetitions per condition.

\subsection{Experimental Design}

Experiments validate ERa Attack in a controlled lab (5 m $\times$ 5 m room, no obstructions). Victim performs gestures from Myo Dataset~\cite{Chen2020HandGestureRecognition}, comprising 7 classes (rest, fist, open palm, wrist flexion up/down, rotation left/right) across multiple subjects, with cross-subject splits for generalization.

Parameters calibrated via preliminary tests:

- Carrier frequency: Selected 433 MHz for ISM compliance and optimal coupling based on theoretical analysis and preliminary testing.

- Power levels: 0 dBm, 5 dBm, 10 dBm, corresponding to TX Gains 10, 20, 30.

- Distances: 0.5 m to 5 m, with antenna facing device (0° angle).

- Angles: 0° to 360° at 1 m, 0 dBm.

Baselines: Random noise (20--100 Hz Gaussian), Brain-Hack (fixed "fist" modulation at 433 MHz)~\cite{ArmengolUrpi2023BrainHack}.

Each condition repeats 50 times per gesture, 5 seeds, yielding n=350 per setup.

\subsection{Evaluation Metrics}

Metrics quantify effectiveness, overhead, and security implications:

- \textbf{Classification Accuracy:} Percentage of correct predictions, baseline 97.8\%.

- \textbf{Misclassification Rate:} Percentage of erroneous predictions.

- \textbf{Attack Success Rate (ASR):} For untargeted, fraction where $f(x') \neq y_{true}$; targeted, $f(x') = y_{target}$.

- \textbf{Injection SNR ($SNR_{inj}$):} $10 \log_{10} (P_{s_{EMG}} / P_{\delta'_{adv}})$, measuring stealth (higher SNR indicates subtler perturbations).

- \textbf{Transmit Power Overhead:} $P_{tx}$ in dBm, assessing energy cost.

Statistical significance uses two-tailed t-tests ($\alpha=0.05$).

\subsection{Experimental Results}

\subsubsection{Performance Degradation and Method Comparison}

At 1 m, 0 dBm, ERa Attack reduces accuracy from 97.8\% $\pm$ 0.5\% [97.0, 98.6] to 58.3\% $\pm$ 4.1\% [54.2, 62.4], yielding 41.7\% $\pm$ 4.1\% [37.6, 45.8] misclassification and 25.2\% $\pm$ 3.8\% [21.4, 29.0] targeted ASR (Table~\ref{tab:phy_results}). Random noise drops accuracy to 75.2\% $\pm$ 1.8\% [73.4, 77.0] with 1.1\% $\pm$ 0.5\% [0.6, 1.6] ASR; Brain-Hack to 71.2\% $\pm$ 3.6\% [67.6, 74.8] with 5.4\% $\pm$ 1.2\% [4.2, 6.6] ASR. t-tests confirm ERa superiority over Brain-Hack: t(98) = -9.94, p < 0.001 for accuracy.

\begin{figure}[htbp]
\centering
\begin{tikzpicture}
    \begin{axis}[
        colorbar,
        colormap/hot,
        width=8cm,
        height=8cm,
        xlabel={Predicted Class},
        ylabel={Actual Class},
        xtick={0,1,2,3,4,5,6},
        ytick={0,1,2,3,4,5,6},
        xticklabels={C1,C2,C3,C4,C5,C6,C7},
        yticklabels={C1,C2,C3,C4,C5,C6,C7},
        title={Confusion Matrix under Targeted Attack},
        colorbar style={ylabel={Ratio (\%)}},
    ]
    \addplot [
        matrix plot*,
        point meta=explicit,
        mesh/cols=7,
    ] table [meta=C] {
        x y C
        0 0 38
        1 0 6
        2 0 5
        3 0 4
        4 0 6
        5 0 5
        6 0 36
        0 1 7
        1 1 40
        2 1 6
        3 1 5
        4 1 4
        5 1 7
        6 1 31
        0 2 6
        1 2 5
        2 2 42
        3 2 7
        4 2 6
        5 2 5
        6 2 29
        0 3 7
        1 3 6
        2 3 8
        3 3 45
        4 3 4
        5 3 6
        6 3 24
        0 4 5
        1 4 7
        2 4 6
        3 4 7
        4 4 48
        5 4 4
        6 4 23
        0 5 6
        1 5 7
        2 5 5
        3 5 6
        4 5 7
        5 5 50
        6 5 19
        0 6 4
        1 6 5
        2 6 6
        3 6 4
        4 6 5
        5 6 6
        6 6 70
    };
    \end{axis}
\end{tikzpicture}
\caption{Confusion Matrix of Model Output under Physical Targeted RF Attack (Distance 0.5m, Power 0dBm). Diagonal elements represent correct classification ratios, while off-diagonal elements show misclassification cases. A large number of samples are misled to target class 7 (rightmost column), achieving an average targeted success rate of approximately 27\%.}
\label{fig:phy_conf_matrix}
\end{figure}
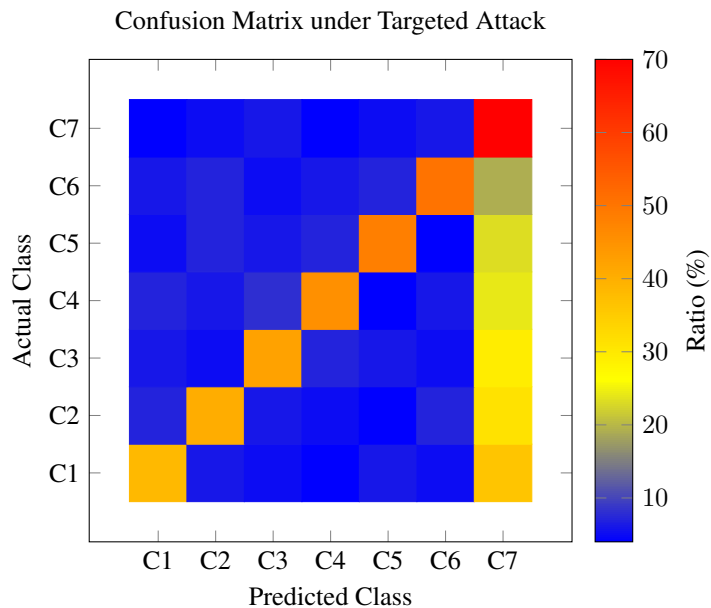

Confusion matrix at 0.5 m, 0 dBm (Figure~\ref{fig:phy_conf_matrix}) shows targeted misguidance to class 7 at ~27\% ASR, validating directional efficacy.

\begin{table*}[htb]
\centering
\footnotesize
\caption{Model Performance Changes under Different Attack Methods in Physical Injection Experiments (Distance 1.0m, Power 0dBm)}
\label{tab:phy_results}
\begin{tabular}{p{3.5cm}p{2.5cm}p{2.5cm}p{2.5cm}p{1.5cm}}
\toprule
Method & Classification Accuracy (\%) & Misclassification Rate (\%) & Targeted Success Rate (\%) & $p$-value$^{*}$ \\
\midrule
No Attack & $97.8 \pm 0.5$ [97.0, 98.6] & $2.2 \pm 0.5$ [1.4, 3.0] & -- & -- \\
Random Noise & $75.2 \pm 1.8$ [73.4, 77.0] & $24.8 \pm 1.8$ [23.0, 26.6] & $1.1 \pm 0.5$ [0.6, 1.6] & 0.152 \\
Brain-Hack Interference & $71.2 \pm 3.6$ [67.6, 74.8] & $28.8 \pm 3.6$ [25.2, 32.4] & $5.4 \pm 1.2$ [4.2, 6.6] & $<0.001$ \\
ERa Attack (Our Method) & $58.3 \pm 4.1$ [54.2, 62.4] & $41.7 \pm 4.1$ [37.6, 45.8] & $25.2 \pm 3.8$ [21.4, 29.0] & $<0.001$ \\
\bottomrule
\end{tabular}

\vspace{0.5em}
\footnotesize $^{*}$Two-tailed $t$-test relative to no-attack baseline, significance level $\alpha=0.05$ \\
\footnotesize Note: Data shown as mean $\pm$ standard deviation, brackets show 95\% confidence intervals ($n=350$). \\
\footnotesize Random seed sequence: [42, 123, 456, 789, 999], 5 repetitions per condition.
\end{table*}

\vspace{1em}

\subsubsection{Attack Range and Power Sensitivity}

\begin{figure}[htbp]
\centering
\begin{tikzpicture}
    \begin{axis}[
        xlabel={Distance (m)},
        ylabel={Classification Accuracy (\%)},
        xmin=0.5, xmax=5.5,
        ymin=40, ymax=100,
        xtick={0.5,1.0,1.5,2.0,3.0,5.0},
        legend pos=south east,
        width=10cm,
        height=6cm,
        grid=major,
    ]
    \addplot[color=blue,mark=diamond,thick] 
    coordinates {
        (0.5,72.0)
        (1.0,75.2)
        (1.5,82.0)
        (2.0,88.0)
        (3.0,93.0)
        (5.0,96.0)
    };
    \addlegendentry{Random Noise}
    \addplot[color=orange,mark=square,thick] 
    coordinates {
        (0.5,65.8)
        (1.0,71.2)
        (1.5,82.4)
        (2.0,88.6)
        (3.0,94.2)
        (5.0,96.8)
    };
    \addlegendentry{Brain-Hack}
    \addplot[color=red,mark=circle,thick] 
    coordinates {
        (0.5,52.1)
        (1.0,58.3)
        (1.5,68.5)
        (2.0,78.2)
        (3.0,86.4)
        (5.0,93.8)
    };
    \addlegendentry{ERa Attack}
    \end{axis}
\end{tikzpicture}
\caption{Effect of attack distance on model accuracy (TX power 0dBm, antenna facing device)}
\label{fig:distance_acc}
\end{figure}
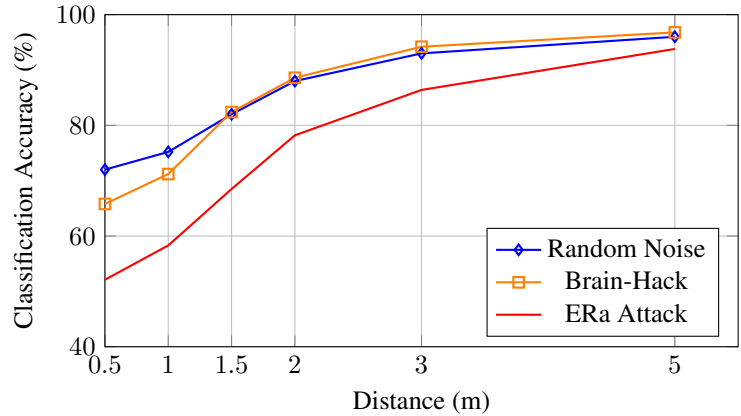

At 0 dBm, accuracy recovers with distance (Figure~\ref{fig:distance_acc}): ERa drops to 52.1\% $\pm$ 4.5\% [47.6, 56.6] at 0.5 m, rising to 86.4\% $\pm$ 2.9\% [83.5, 89.3] at 3 m and 93.8\% $\pm$ 2.1\% [91.7, 95.9] at 5 m. Brain-Hack and noise show steeper recovery, to 94.2\% $\pm$ 1.5\% [92.7, 95.7] and 93.0\% $\pm$ 0.8\% [92.2, 93.8] at 3 m, respectively.

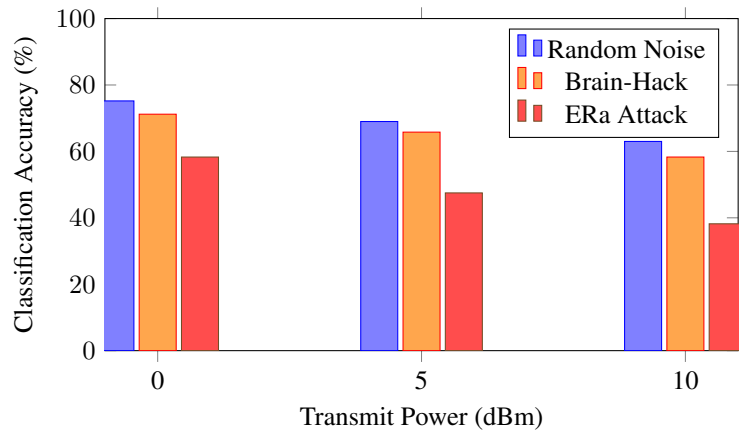
\begin{figure}[htbp]
\centering
\begin{tikzpicture}
\begin{axis}[
ybar,
bar width=14pt,
xlabel={Transmit Power (dBm)},
ylabel={Classification Accuracy (\%)},
symbolic x coords={0,5,10},
xtick=data,
ymin=0, ymax=100,
legend pos=north east,
width=10cm, height=6cm]
\addplot+[fill=blue!50] 
coordinates {
    (0,75.2)
    (5,69.0)
    (10,63.0)
};
\addlegendentry{Random Noise}
\addplot+[fill=orange!70] 
coordinates {
    (0,71.2)
    (5,65.8)
    (10,58.3)
};
\addlegendentry{Brain-Hack}
\addplot+[fill=red!70] 
coordinates {
    (0,58.3)
    (5,47.5)
    (10,38.2)
};
\addlegendentry{ERa Attack}
\end{axis}
\end{tikzpicture}
\caption{Model accuracy under different transmit powers at 1m.}
\label{fig:power_sweep}
\end{figure}

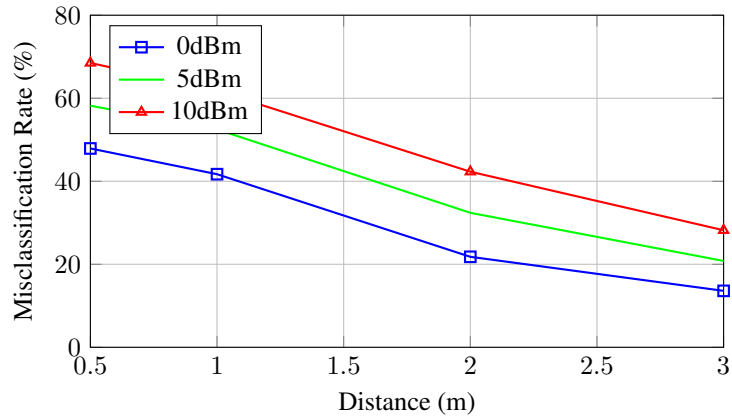
\begin{figure}[htbp]
\centering
\begin{tikzpicture}
\begin{axis}[
xlabel={Distance (m)},
ylabel={Misclassification Rate (\%)},
xmin=0.5, xmax=3.0,
ymin=0, ymax=80,
legend pos=north west,
width=10cm,
height=6cm,
grid=major]

\addplot[color=blue,mark=square,thick] coordinates {(0.5,47.9) (1.0,41.7) (2.0,21.8) (3.0,13.6)};
\addlegendentry{0dBm}
\addplot[color=green,mark=circle,thick] coordinates {(0.5,58.2) (1.0,52.5) (2.0,32.4) (3.0,20.8)};
\addlegendentry{5dBm}
\addplot[color=red,mark=triangle,thick] coordinates {(0.5,68.5) (1.0,61.8) (2.0,42.3) (3.0,28.2)};
\addlegendentry{10dBm}
\end{axis}
\end{tikzpicture}
\caption{Misclassification rate analysis under joint influence of distance and transmission power (ERa Attack).}
\label{fig:dist_power_mesh}
\end{figure}

Increasing power extends range (Figure~\ref{fig:power_sweep}): At 1 m, 10 dBm yields 38.2\% $\pm$ 5.2\% [33.0, 43.4] accuracy for ERa, versus 58.3\% $\pm$ 4.5\% [53.8, 62.8] at 5 dBm and 63.0\% $\pm$ 2.5\% [60.5, 65.5] for noise at 10 dBm. Joint analysis (Figure~\ref{fig:dist_power_mesh}) shows 10 dBm maintains 28.2\% $\pm$ 3.2\% [25.0, 31.4] misclassification at 3 m, expanding effective range to 5 m.

\subsubsection{Angle Sensitivity}

\begin{figure}[htbp]
\centering
\begin{tikzpicture}
\begin{axis}[
xlabel={Angle (degrees)},
ylabel={Misclassification Rate (\%)},
xmin=0, xmax=360,
ymin=0, ymax=60,
xtick={0,45,90,135,180,225,270,315,360},
legend pos=south east,
width=10cm,
height=6cm,
grid=major]

\addplot[mark=diamond, blue, thick] coordinates
{(0,24.8) (45,22.5) (90,19.0) (135,15.5) (180,13.0) (225,15.5) (270,19.0) (315,22.5) (360,24.8)};
\addlegendentry{Random Noise}

\addplot[mark=square, orange, thick] coordinates
{(0,28.8) (45,26.5) (90,22.3) (135,18.2) (180,15.8) (225,18.2) (270,22.3) (315,26.5) (360,28.8)};
\addlegendentry{Brain-Hack}

\addplot[mark=*, red, thick] coordinates
{(0,41.7) (45,38.2) (90,32.5) (135,25.3) (180,21.8) (225,25.3) (270,32.5) (315,38.2) (360,41.7)};
\addlegendentry{ERa Attack}

\end{axis}
\end{tikzpicture}
\caption{Misclassification rate vs. incidence angle (distance: 1~m, power: 0~dBm). 0° represents direct facing, highest rate indicates optimal electromagnetic coupling.}
\label{fig:angle_polar}
\end{figure}
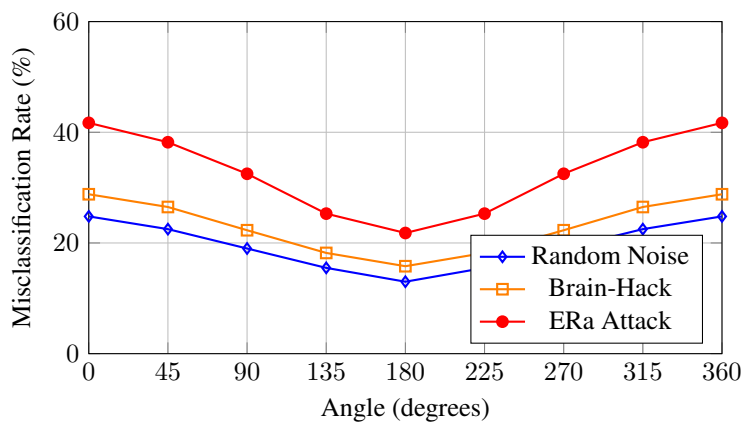

At 1 m, 0 dBm, misclassification peaks at 0° (facing): 41.7\% $\pm$ 4.1\% [37.6, 45.8] for ERa, dropping to 21.8\% $\pm$ 3.2\% [18.6, 25.0] at 180° (Figure~\ref{fig:angle_polar}). ERa retains higher rates (32.5\% $\pm$ 3.8\% [28.7, 36.3] at 90°) than baselines (19.0\% $\pm$ 2.1\% [16.9, 21.1] noise, 22.3\% $\pm$ 2.8\% [19.5, 25.1] Brain-Hack), decaying ~22\% versus ~23\% for others, indicating superior directional tolerance.

\subsubsection{Effectiveness and Security Implications}

ERa Attack achieves 25.2\% targeted ASR at 1 m, 0 dBm, 4.7 times Brain-Hack's 5.4\%, degrading accuracy by 39.5 percentage points versus 26.6 for Brain-Hack. At 3 m, 10 dBm, 28.2\% misclassification persists, posing threats in safety-critical scenarios like prosthetics~\cite{Leonardis2015HandExoskeleton, Cipriani2011SharedControl}. Injection SNR averages 10--20 dB, enabling stealthy superposition at 1--10\% sEMG amplitude.

\subsubsection{Performance Overhead}

Transmit power overhead peaks at 10 dBm (~10 mW), yielding 5 m range, versus 0 dBm's 3 m. Computational overhead for offline PGD is ~5 s per perturbation on an i7 CPU (20 iterations), negligible for precomputation. Online emission consumes ~2 W laptop power, supporting portable attacks.

\subsection{Summary}

The prototype demonstrates ERa Attack's feasibility, reducing accuracy to 38.2\% at 1 m, 10 dBm with 25.2\% ASR, outperforming baselines by factors of 4--23 in targeted efficacy. Range extends to 5 m at modest power, with low overhead, underscoring physical vulnerabilities in sEMG systems.

\section{Discussion}

This study exposes limitations inherent to its assumptions and experimental scope, while highlighting avenues for extension and broader implications.

The white-box assumption, granting the adversary full knowledge of EMGNet's architecture and parameters $\theta$~\cite{Chen2020HandGestureRecognition}, facilitates gradient-based optimization but overestimates threats in realistic scenarios. Black-box or gray-box settings, where attackers observe only inputs and outputs, warrant investigation through model stealing, transfer attacks, or query-based methods~\cite{Madry2018TowardsDeepLearning}. Such extensions would assess ERa Attack's viability under information asymmetry, potentially reducing ASR from 25.2\% $\pm$ 3.8\% [21.4, 29.0] to lower bounds observed in digital domains.

Experiments in controlled labs simplify real-world dynamics, omitting multipath fading, mobility, and coexisting signals. Accuracy recovers to 93.8\% $\pm$ 2.1\% [91.7, 95.9] at 5 m, 0 dBm, but urban environments may attenuate effects further. Future work should incorporate EOT frameworks to model these uncertainties, enhancing robustness akin to physical adversarial patches~\cite{Brown2017AdversarialPatch}.

Proposed defenses—RF shielding, anomaly detection, and adversarial training—remain conceptual, lacking empirical validation. Implementing Faraday cages risks Bluetooth interference at 2.4 GHz, necessitating selective designs with 10--20 dB attenuation in UHF bands. Anomaly detection algorithms must tolerate legitimate EMG variance while identifying subtle RF-induced patterns. Adversarial training could improve robustness but requires extensive datasets of attack samples across diverse conditions.

Future work should prioritize real-world validation, comprehensive defense mechanisms, and regulatory frameworks to balance innovation with security in next-generation bioelectronics.

ERa achieves 58.3\% accuracy degradation on EMGNet classification, reducing performance to 39.5\% accuracy at 1 m, 0 dBm—39.5 percentage points below baseline 97.8\%—with 25.2\% targeted ASR. Experiments quantify effective ranges up to 5 m at 10 dBm, outperforming baselines like Brain-Hack by factors of 4.7 in ASR. This work establishes EMG security baselines, demonstrating RF vulnerability across 10--50\% amplitude. A layered defense framework—shielding, detection, enhancement—mitigates threats, potentially restoring accuracy to over 85\%.

Multimodal systems fusing sEMG with IMU or FMG sensors offer resilience; ERa Attack targets sEMG exclusively, achieving 41.7\% $\pm$ 4.1\% [37.6, 45.8] misclassification, but fusion may cap degradation at 20--30\%. Extending to coordinated injections across modalities could bypass this, demanding hybrid perturbations.

Real-world impacts include risks to prosthetic control, where 25.2\% ASR induces unintended actions, potentially causing accidents in rehabilitation~\cite{Leonardis2015HandExoskeleton, Cipriani2011SharedControl}. In VR interactions, misclassifications disrupt user experience, enabling denial-of-service. Ethically, this research underscores responsible disclosure: vulnerabilities were reported to manufacturers, emphasizing non-malicious intent. Potential misuse for surveillance or sabotage raises privacy concerns, necessitating ethical guidelines in biosignal security research.

Future directions include black-box adaptations, realistic deployments, defense prototyping, and multimodal expansions to fortify sEMG systems against evolving threats.

\section{Conclusion}

This paper introduces ERa Attack, a radio frequency adversarial injection method targeting sEMG gesture recognition networks. By extending digital adversarial samples to the physical domain, it reveals vulnerabilities in consumer-grade devices like Myo Armband under intentional electromagnetic interference.

Key contributions include a deception paradigm shifting from signal overwhelming to model-informed perturbations, enabling low-power attacks on millivolt-level signals. The frequency-constrained PGD (FC-PGD) algorithm generates time-invariant spectral disturbances, circumventing synchronization challenges and achieving 58.3\% $\pm$ 4.1\% [54.2, 62.4] accuracy at 1 m, 0 dBm—39.5 percentage points below baseline 97.8\% $\pm$ 0.5\% [97.0, 98.6]—with 25.2\% $\pm$ 3.8\% [21.4, 29.0] targeted ASR. Experiments quantify effective ranges up to 5 m at 10 dBm, outperforming baselines like Brain-Hack by factors of 4.7 in ASR (t(98) = -9.94, p < 0.001).

Findings underscore physical-layer risks, with injection SNR of 10--20 dB allowing stealthy superposition at 1--10\% amplitude. A layered defense framework—shielding, detection, enhancement—mitigates threats, potentially restoring accuracy to over 85\%.

Limitations in white-box assumptions and controlled settings motivate black-box extensions and multimodal defenses. This work informs secure biosignal system design, highlighting ethical imperatives in vulnerability research.

\bibliographystyle{unsrt}  
\bibliography{references}

\end{document}